\documentclass{acm_proc_article-sp}
\usepackage{url}
\usepackage{amsmath}
\usepackage{amsfonts}
\usepackage{algorithmic}
\usepackage{algorithm}
\DeclareMathOperator*{\argmax}{argmax}

\begin{document}

\title{Extracting Geospatial Preferences Using\\ Relational Neighbors}

%
%
%
%

\numberofauthors{6} 
\author{
%
\alignauthor
Leandro Balby Marinho\\
        \affaddr{\normalsize Federal University of Campina Grande, Brazil}\\
       \email{\normalsize lbmarinho@dsc.ufcg.edu.br}
\alignauthor
Cl\'{a}udio de Souza Baptista\\
        \affaddr{\normalsize Federal University of Campina Grande, Brazil}\\
       \email{\normalsize baptista@dsc.ufcg.edu.br}
\alignauthor Thomas Sandholm\\
        \affaddr{\normalsize HP Labs, USA}\\
       \email{\normalsize thomas.e.sandholm@hp.com}
\and  
\alignauthor Iury Nunes\\
        \affaddr{\normalsize Federal University of Campina Grande, Brazil}\\
       \email{\normalsize iury.nunes@ccc.ufcg.edu.br}
\alignauthor Caio N\'{o}brega\\
        \affaddr{\normalsize Federal University of Campina Grande, Brazil}\\
       \email{\normalsize caio.nobrega@ccc.ufcg.edu.br}
\alignauthor Jord\~{a}o Ara\'{u}jo\\
        \affaddr{\normalsize Federal University of Campina Grande, Brazil}\\
       \email{\normalsize jordao.araujo@ccc.ufcg.edu.br}
}

\maketitle
\begin{abstract}
With the increasing popularity of location-based social media applications and devices that automatically tag generated content with locations, large repositories of collaborative geo-referenced data are appearing on-line. Efficiently extracting user preferences from these data to determine what information to recommend is challenging because of the sheer volume of data as well as the frequency of updates. Traditional recommender systems focus on the interplay between users and items, but ignore contextual parameters such as location. In this paper we take a geospatial approach to determine locational preferences and similarities between users.
We propose to capture the geographic context of user preferences for items using a relational graph, through which we are able to derive many new and state-of-the-art recommendation algorithms, including combinations of them, requiring changes only in the definition of the edge weights. Furthermore, we discuss several solutions for cold-start scenarios. Finally, we conduct experiments using two real-world datasets and provide empirical evidence that many of the proposed algorithms outperform existing location-aware recommender algorithms.
\end{abstract}



\section{Introduction}\label{sec:introduction}
With the affordable prices of GPS-enabled mobile devices and the success of social networks, location-based social media has  become increasingly popular in recent years. Users can upload content, e.g., photos, videos, and text, and annotate that content with geographical identification metadata, typically known as \textit{geotags}. Geotags act as geographic indexes helping users to organize and retrieve location-specific information. Foursquare\footnote{\url{http://foursquare.com/}}, as an example, is a location-based service where users endorse and share tips about visited points of interest (POI). It reached 725 thousand registered users and 22 million check-ins (i.e. endorsed POIs) in 2010\footnote{\url{http://mashable.com/2010/03/29/foursquare-growth-numbers/}}. 


Recommender systems (RS) are among the best known techniques for helping users filter out and discover relevant information in large data sets. In the typical scenario, RS algorithms exploit user-item matrices representing user preferences for items, e.g.,  the rating history of purchased books in Amazon, with the aim of recommending the items most likely to be relevant to the user. While most of the RS work to date has ignored the locations where users demonstrated interest for an item, there are many scenarios in which the geographic context of an item has a direct influence on the preferences of the user for that item. Mao et al.~\cite{ye2011}, for example, showed that Foursquare users prefer POIs that are nearby the POIs they already visited in the past, while \cite{matyas2009} showed that Panoramio users who took pictures in nearby locations in the past, tend to share similar preferences of geographic regions in the future. Efficient extraction and representation of location-specific user preferences are thus essential to decide on what item to recommend.

Location-aware recommender systems suggest relevant geotagged items for a given user within a declared geographic area. Relevance here can assume different notions, depending on the geographic constraints imposed by the user. For example, a user may be interested in objects nearby his previous, current, or future location within a given radius. For example, a first time user visiting the Stanford University campus in Palo Alto, US, might be interested to know what is worth visiting inside the campus, while a second time visitor may want to know what else is worth visiting. Each of these scenarios can lead to various definitions of user preference, hence it is important to know which definition works best for each scenario. The literature concerning location-aware recommender system is still sparse, where the methods are ad-hoc and can not be easily changed to meet the different recommendation scenarios outlined above (e.g.~\cite{matyas2009,sandholm2011,prete2010}).

In this paper we propose a relational graph for capturing the geographic context of users that suits all the aforementioned recommendation scenarios. We introduce several strategies to represent location-specific user preferences in the graph and show how to derive many recommendation algorithms, including ensembles of them, by only changing the definition of the edge weights. Our contributions are as follows:

\begin{enumerate}
 \item We propose a new model for geotagged data that is able to capture both the geographic context of users and their preferences for objects within the declared geographic context.
 \item We introduce new similarity measures that take into account the spatial decisions of users.
 \item We propose a recommendation algorithm based on a relational neighbor graph, that derive many recommendation algorithms and ensembles of them, as special cases, by only requiring changes in the definition of the edge weights.
 \item Finally, we conduct experiments using two datasets in various recommendation scenarios, including cold-start ones, and provide empirical evidence that many of the proposed algorithms outperform existing location-aware recommender algorithms.
\end{enumerate}

The rest of the paper is organized as follows. Section~\ref{sec:problem} introduces the problem setting. Section~\ref{sec:approach} presents our relational graph representation of the data, weighting schemes for capturing user geographic preferences, and a recommendation algorithm based on relational neighbors. Section~\ref{sec:experiments} presents the experimental setting and evaluation. Section~\ref{sec:related_work} describes related work, and Section~\ref{sec:conclusions} concludes the paper and discusses future work.

\section{Problem Setting}
\label{sec:problem}

The recommendation scenario is as follows. A user specifies the geographic region of interest, e.g., a city or a region in that city, and the recommender engine suggests items within the declared geographic region that are likely to be relevant to the user. 

So, let $U$ be the set of users, $G$ the set of regions denoting geographic contexts, and $I$ the set of geotagged items. Notice that what is meant by a region is application dependent since regions can assume different geographic shapes, such as point coordinates, circles, lines, and polygons. In this paper we only consider implicit feedback data\footnote{Although our framework can trivially support explicit feedback as well.}, i.e., the set $S\subseteq U\times G\times I$ of ternary relations between users, geographic contexts, and geotagged items. The task is then to find a prediction scoring function
\begin{equation}
\hat{s}:U\times G\times I\rightarrow \mathbb{R}
\label{eq:score}
\end{equation}
that predicts a preference score for items within certain geographic regions, given a target user. Now, for a given user $u\in U$, and a given geographic context $g\in G$, the topN recommendations can be computed by
\begin{equation}
\textit{topN}(u,g):=\argmax_{i\in I_g}^n \hat{s}(u,g,i)
\label{eq:topn}
\end{equation}
where $n$ denotes the topN items to be recommended and $I_g$ the set of items within the geographic context $g$. For convenience, we also define $I_{u,g}:=S\cap(\{u\} \times \{g\} \times I)$ as the set of items of user $u\in U$ in a given geographic context $g\in G$.

\section{A Relational Approach}
\label{sec:approach}

Relational classification refers to an active area of machine learning where classifiers usually consider, additionally to the typical attribute-value data of objects, relational information. A scientific paper, for example, can be connected to another paper that has been written by the same author or because they share common citations. It has been shown that in many classification problems, relational classifiers outperform purely attribute-based classifiers~\cite{chak98,Lu03link-basedclassification,preisach06ensemble}. In particular, Macskassy and Provost~\cite{Macskassy03asimple} showed that simple relational neighbor-based techniques, besides requiring low computational costs, perform competitively to, and in some cases even outperforms, more complex relational methods such as Probabilistic Relational Models and Relational Probability Trees. The basic idea is that the classification of a target instance solely depends on the class labels of related instances of the same type. Since geotagging data is inherently relational, we propose to capture the geographic context of users and items by a relational graph, which, as a side effect, gives us many tools from relational classification that can be directly applied to the location-aware recommendation problem. 

In order to easily use neighborhood-based classification methods, similarly to~\cite{Macskassy03asimple,preisach06ensemble}, we adopt a homogenous view of the relations in the data. In a homogeneous view we have only one entity type, such that, there is a set of target entities $x\in X$ and relations $R\subseteq X\times X$ between them. However, as we saw in Section~\ref{sec:problem}, geotagging data forms a set $S$ of ternary relations between three different types of entities. Therefore, we first need to convert these ternary relations into the desired homogeneous relations. We do this as follows~\footnote{relational classification is possible with heterogenous node types as well
as has been demostrated in~\cite{jeh2002} and~\cite{minkov2010} but the added complexity defeats our goal of fast model building and evaluation.}.

\subsection{Graph Definition}
\label{sec:graph}

First let $V:=\{(u,g)\,|\,\exists i\in I\,:\,(u,g,i)\in S\}$ denote the set of all distinct user/geographic context combinations in $S$. We now propose to interpret $S$ as a graph $\mathbb{G}:=(V,E)$, where $V$ is the set of vertices and $E\subseteq V\times V$ the set of edges. We assume that there is an edge between two vertices if they share the same geographic context, i.e., $$\{(v,v')\in E \,|\, g_v = g_{v'}\}$$ For convenience, let $g_v:=g$ and $u_v:=u$ denote the geographic context and the user of node $v\in V$ respectively. In other words, we assume that users who share the same geographic region are related to some extent.  

\begin{figure}
\centering
\includegraphics[width=0.5\linewidth]{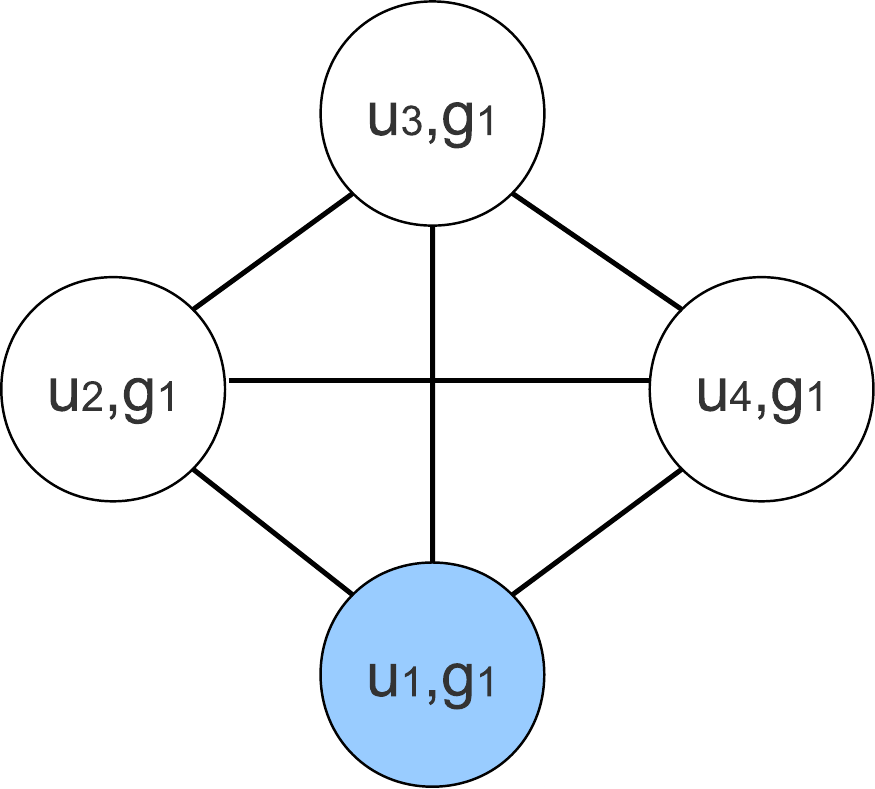}
\caption{Example of relational graph for capturing geographic context.}\label{fig:graph}
\end{figure}

Now suppose that John is a first time visitor in Rio de Janeiro, Brazil, and wants to know from other people who already have been in Rio which places are worth visiting there.
To continue our example, we denote the pair John/Rio by a colored node in the graph of Figure~\ref{fig:graph}, i.e., 
the target node for which we want to compute recommendations. 
John is denoted as $u_1$ and Rio as $g_1$ respectively, and the other nodes connected to it contain the users who already have been in Rio. For computing recommendations, we just need to go through the items of the neighbors and define some selection criterion on which items to recommend. 

This idea assumes that entities related to each other, in this case users sharing the same geographic context, 
are similar and tend to select the same items. Notice that the strength of the similarity depends on the size of the geographic region being considered. If the declared geographic context is a small region, say the Copacabana beach in Rio de Janeiro, then it is more likely that users within this area will be more strongly related to each other than users sharing larger geographic contexts, such as the whole country of Brazil. We can alleviate these effects by defining appropriate weights to each relation, as we will see next.

\subsection{Weighting Schemes}
\label{sec:weight}
In this section we present several weighting strategies for the edges of the relational graph. And as we will see in the next section, each weighting scheme leads to a specific recommendation algorithm, without, however, the need to change the overall algorithm.

\paragraph{Uniform Weighting}

Here the same weight value is assigned to each edge ($1$ in our case), denoting that each neighbor is of the same importance to the target user. Therefore, the weight $w$ for any edge $(v,'v)\in E$, is defined as $w^{\text{uni}}(v,v'):=1$. The computational cost for assigning uniform weights to the relations of a target node $v$ is in the order of $O(|N_v|)$, i.e., a linear scan in the neighbors of $v$, here denoted by $N_v$.

\paragraph{Correlation Weighting} For two nodes $(v,v')\in E$, we can represent $v$ and $v'$ as profile vectors, where each component of the vector is a geotagged item and the values denote the preference of a user for an item, i.e., $$\vec{m}_v:=(i_1,...,i_{|I|})$$ We can either construct this vector solely based on the items of the geographic context of interest, or we could consider the items of all geographic contexts. The assumption is that users who have selected the same items within the same geographic contexts are more similar than users who did not. If a user declared interest for some geographic context for which he has not yet selected any item, it will not be possible to compute any correlation similarity with other users, unless we build the profile vectors considering the items of other geographic contexts.

The edge weight between two nodes is finally computed by applying a correlation metric between the desired nodes' profile vectors, which in our case, is the cosine similarity: 
\begin{equation}
w^{\text{cor}}(v,v'):=
      \frac{\langle \vec{m}_v,\vec{m}_{v'} \rangle}{\|\vec{m}_v\| \|\vec{m}_{v'}\|}\label{eq:cosine}
\end{equation}

The computational cost for assigning cosine similarities to the relations of a target node $v$ is in the order of $O(Z\cdot |N_v|)$ since we need to compute $|N_v|$ similarities, each requiring $Z$ operations.

\begin{figure}
\centering
\includegraphics[width=0.55\linewidth]{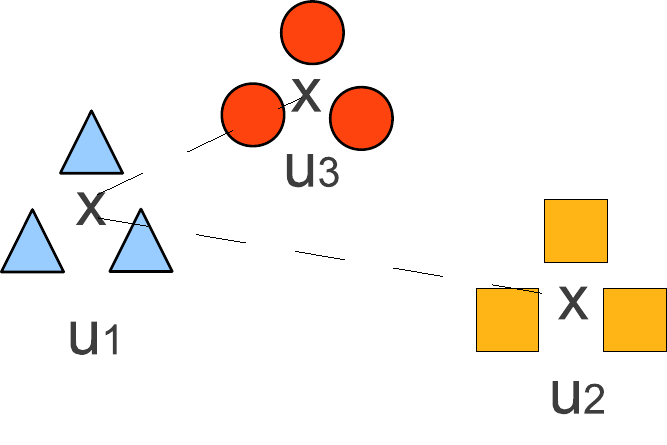}
\caption{Example of geographic similarity. The items of users $u_1$, $u_2$, $u_3$ are represented by triangles, circles and rectangles respectively. Centroids are represented by ``X'' and distances by dashed lines. Thus, $u_3$ would be regarded as more similar to $u_1$ than $u_2$. }\label{fig:geo_sim}
\end{figure}

\paragraph{Geographic Similarity} We can use the geographic distances between users' items to define the strength of their relation. The assumption is that users who select items nearby the items of other users should be closely related. Therefore, for defining the weight between two nodes $(v,v')\in E$ through geographic similarity, we first calculate the geographic centroids of the set of items of users $u_v$ and $u_{v'}$. For computing the geographic centroid of a given user $u\in U$, we sum up the coordinates of each geotagged item of user $u$ within the declared geographic context $g$, i.e., the items in the set $I_{u,g}$, and divide the resulting sum by the number of items in the set
\begin{align*}
\frac{1}{|I_{u,g}|}\displaystyle\sum_{i\in I_{u,g}} p(i)\\
\end{align*}
where $p(i)$ returns the latitude/longitude coordinate used to geotag item $i\in I$. Now, the geographic distance between $u(v)$ and $u(v')$ centroids is given by any of the many existing functions for calculating geodetic distances between latitude/longitude
coordinates, e.g. the Haversine formula. Here we denote such a function by  $d(x,y)$ where $x$ and $y$ are two coordinates. Finally, we only need to turn the distance into a similarity and bound it to the range $[0,1]$, which is done as follows
\begin{equation}\label{eq:geo_sim}
	w^{\text{geo}}(v,v') := 1-\frac{d\left (c_{u_{v}},c_{u_{v'}}\right )}{d_{\text{max}}} 
\end{equation}
where $d_{max}$ is the maximal possible distance between any points in the region of interest and $c_u$ is the geographic centroid of the set of items of user $u$. In other words, when the distance between the items of two users is small, their similarity tend to $1$, and is $0$ when the distance equals $d_{\text{max}}$. Figure~\ref{fig:geo_sim} illustrates this idea. Noticed that for this to work, we are assuming that users tend to form geographic clusters among the selected items. This was empirically observed in~\cite{ye2011} by showing that Foursquare users tend to check-in to POIs that are nearby the POIs they have already visited. 

The computational cost of this weighting scheme is in the same order as the \textit{correlation weighting} above.
 
\paragraph{Partonomy-Based Similarity} The different granularities between geographic contexts can be represented by a hierarchical \textit{partonomy}, i.e., a graph $\mathbb{G}:=(G,E)$ where vertices are represented by geographic contexts and each edge $e\in E$ represents a part-of relation between two geographic regions~\cite{matyas2009}. For example, a POI is a part of a city, a city is a part of a state, a state is a part of a country, etc. Matyas et al.~\cite{matyas2009} proposed several similarity measures for capturing user similarities with respect to specific levels of a weighted geographic partonomy. The idea is that if two users are not found to be similar in a lower level of the hierarchy, they may, eventually, be found to be similar in a higher level. For example, even if two users did not visit the same places in some city, they still can be considered to be similar since they have visited the same city. We can use these measures, together with a partonomy, to weigh the relations in our relational graph. Here we just present the best performing measure reported in~\cite{matyas2009}.

The similarity of two users $u,u'\in U$ with respect to node $g$ is calculated as follows

\begin{equation}
 \textit{sim}^{\textit{inf}}(u,u',g):=\frac{\sum_{(c\in A_g\cap B_g)}\textit{information}(c)}{\sum_{(c\in A_g\cup B_g)}\textit{information}(c)}
\end{equation}
where $A_g$ and $B_g$ denote the sets of children nodes of $g$ in which user $u$ and $u'$ selected items respectively. The similarity is weighted by the function \textit{information}, which can be seen as the inverse of the popularity of a node. The assumption is that users sharing less popular nodes are more similar than users sharing more popular ones~\cite{matyas2009}.

Now, for computing the similarity of two users $u,u'\in U$ with respect to a certain partonomy layer $l$, the following formula is used
\begin{equation}\label{eq:two-layer}
 \textit{sim}^{\textit{two-layer}}(u,u'):=\frac{\sum_{(g\in G_l)}\textit{sim}^{\textit{inf}}(u,u',g)\cdot w^{\textit{node}}(g)}{\sum_{(g\in G_l)}w(g)}
\end{equation}
where $G_l$ is the set of nodes in the $l$-th layer of the partonomy. This measure was named \textit{two-layer similarity} in~\cite{matyas2009} because when computing the similarity of two users in a certain layer, the measure uses the layer immediately below.

Returning the discussion to the weighting of the edges in our relational graph, we can now define the weight of a certain edge $(v,v')\in E$ in terms of the \textit{two-layer similarity}, i.e., 
\begin{equation}\label{eq:weight-two-layer}
w(v,v'):=\textit{sim}^{\textit{two-layer}}(u_v,u_{v'})
\end{equation}

This similarity can be quite expensive to compute, given all the necessary steps described above for weighting the partonomy. Even assuming that the weighted partonomy is given, the complexity for computing the weight between $v$ and all its neighbors is still higher than for the other similarities. The computational cost is in the order of $$O(|N_v||G_l||C_l|\cdot Z)$$ where $C_l$ is the set of children nodes of the $l$-th layer, since for each neighbor, we need to go through all the locations $g$ in the $l$-th layer of the partonomy, for each location go through all its children, and finally for each child we need to perform $Z$ operations.


\subsection{Recommendation Algorithm}
\label{sec:computing}
 Now, we have all the components for deriving a location-aware recommendation procedure. Algorithm~\ref{algo:location} describes the overall recommendation process. It receives as input a weighted graph $\mathbb{G}$, a user/geographic context pair denoted by $v$, for which we want to generate recommendations, and the number $n$ of recommendations to be returned. The algorithm iterates through the neighbors of $v$ (line 3), denoted by $N_v$, and for each neighbor, it iterates through the items within the geographic context of interest (line 4) and accumulates weights in the array \textit{scores}, that is indexed by items. Finally, it sorts the scores in descending order of weights and presents the top-$n$ geotagged items that the target user has not already selected. 

Note that if we define all weights to 1, we end up recommending the most popular items within the geographic context of interest. Or if we decide to weigh edges according to the similarities presented in Section~\ref{sec:weight}, we end up with many flavors of collaborative filtering-based algorithms. 

Assuming that the weighted relational graph is given, the complexity of this algorithm only depends on the computation of a weighted sum of geotagged items, which means $|N_v|$ passes in the set of items $I$. Hence, the complexity is given by $O\left(|I||N_v|\right)$.
\clearpage

\begin{algorithm}
\caption{Graph-based Location Recommendations}
\begin{algorithmic}[1]
\STATE \textbf{Input:} $\mathbb{G}(V,E)$, $v\in V$, $n$ \\
\STATE \textbf{Output:} list of topN recommendations\\

\FORALL {$v' \in N_v$}
\FORALL {$i \in I_{u_{v'},g_{v'}}$}
\STATE $\textit{scores}[i]\leftarrow \textit{scores}[i] + w(v,v')$ \\
\ENDFOR
\ENDFOR
\STATE $\text{topN}\leftarrow \displaystyle\argmax_{i\in I\setminus I_{u_{v},g_{v}}}^n \textit{scores}[i] $\\

\end{algorithmic}
\label{algo:location}
\end{algorithm}

\section{Experiments}
\label{sec:experiments}
In this section we describe our two datasets, geotagged photos from Panoramio, and print jobs from the HP ePrint Mobile Print Location
service; the evaluation protocol adopted; and the results for each dataset. We considered three different recommendation scenarios.

In the first scenario, we hide all the geotagged items of each test user in a geographic context, 
and use the remaining data for trying to predict the removed items. 
This corresponds to a cold-start scenario where a user has not selected any item in the geographic context of interest. 
We will refer to this scenario as \textit{leave-all-out}.

In the second scenario, we remove some geotagged items for each test user, 4 photos in Panoramio and 1 printer provider in ePrint, and use the remaining data for predicting the removed items. This scenario represents those users who already selected some items in a given location but want to know what other items are worth selecting in this location. We will refer to this scenario as \textit{leave-some-out}

In the third scenario, we have a mix of both scenarios, i.e., some fraction of the users are first time users and the other fraction already have selected some items in the location of interest. This corresponds to a more realistic scenario, and to the best of our knowledge, this is the first time location-aware algorithms are evaluated in this kind of scenario. We will refer to this scenario as \textit{leave-some/all-out}

\subsection{Panoramio Experiments}
\label{sec:panoramio}
Panoramio is a photo-sharing website from Google where users can upload, geotag, and retrieve photos of landmarks. Each photo in Panoramio is georeferenced using latitude and longitude information. Similarly to~\cite{matyas2009}, we assume that if a user takes a picture in a specific location, then he has some interest in that location. As geographic context of interest we have chosen the city of Rio de Janeiro, which is one of the top touristic places in Brazil, so, it has a large set of photos. 

\paragraph{Data Collection and Preparation} In order to retrieve the set of photos taken in Rio, we did a spatial search for photos inside the bounding box of Rio using the data access API of Panoramio. We then iterated through each photo in the result set, and retrieved the users who took
these photos. For each of these users, we then retrieved the other locations where they took photos, as well as the users and photos in those locations. Then we removed the cities with too little activity from users who also visited Rio. Approximately the top three cities in all the crawled states were kept in the
evaluation set. We used the gazetteer of HP Gloe\footnote{\url{http://www.hpgloe.com/}} to obtain the place names. 

In order to use the \textit{two-layer-similarity} weighting scheme presented in Section~\ref{sec:weight}, we built a geographic partonomy, and similarly to~\cite{matyas2009}, we worked with three countries (Brazil, Chile, and USA) as high level nodes, states and cities as intermediary nodes, and geographic clusters as leaf nodes. Our data set contains 35,920 photos (4,906 from Rio) and 7,048 users in total.  

In Panoramio the geographic items are represented by latitude and longitude coordinates, hence, it is very difficult for two users to take a picture in the exact same location. 
Therefore, we adopted the same approach as in~\cite{matyas2009}, where the authors used geographic clusters to represent geographic items. 
So, instead of recommending individual point coordinates, we recommend regions where users may be interested in taking photos. 
For computing the clusters, we used the DBSCAN~\cite{Ester96adensity-based} algorithm with
the following parameters: $\textit{MaxRadius}=1$ Km and $\textit{MinPoints}=3$. The $\textit{MaxRadius}$ parameter was set to 1 Km, 
as we assume that a radius of 1 Km from a given photo is sufficient to establish a geographic similarity between photos, 
e.g., photos taken at the University of Stanford campus. We tested to set $\textit{MaxRadius}=500$ meters without any significant differences in performance, so we do not show those results here. 
Also, the minimum number of points to form a cluster was set to 3, in order to establish popularity of a given POI. We computed 1,187 clusters, 221 of which are in Rio. 

\paragraph{Evaluation Protocol} For testing the algorithms, we considered the dense part of the data, i.e., only the users who have taken at least 5 photos in Rio (see Table~\ref{tab:dataset}). For the \textit{leave-some-out} and \textit{leave-some/all-out} scenarios, we generated 5 random splits of training/test sets and took the average precision and recall on top-$10$ recommendation lists over all splits. We have a hit every time a hidden photo is found to belong to some of the recommended geographic clusters. Whenever CF is not able to fill the recommendation list up to 10, we fill up the list with the most popular items that are not already in the recommendation list.

\begin{table}[t]
\centering
\caption{Panoramio data for Rio de Janeiro}\label{tab:dataset}
\begin{tabular}{|r|r|r|}
 \hline
& $|U|$ & $|I|$ \\
\hline
Training & 1,062 & 4,906 \\
\hline
\hline
Test & 186 & 3,590  \\
\hline

\end{tabular}
 
\end{table}

\begin{figure}[htp]
\vspace{-0.2cm}
\centering
\includegraphics[width=0.99\linewidth]{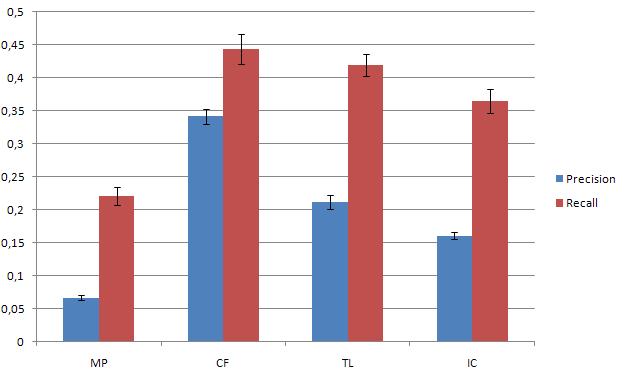}
\caption{Evaluation on \textit{leave-some-out} scenario.}\label{fig:some_out}
\end{figure}

\begin{figure}[htp]
\vspace{-0.2cm}
\centering
\includegraphics[width=0.99\linewidth]{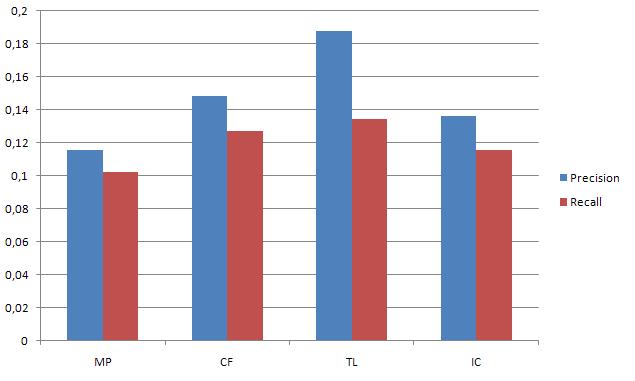}
\caption{Evaluation on \textit{leave-all-out} scenario.}\label{fig:all_out}
\end{figure}

\paragraph{Algorithms} We have used several weighting schemes for our relational graph, which resulted in the following recommendation algorithms:

\begin{itemize}
 \item \textbf{Most popular (MP):} Recommends the most popular geographic clusters in Rio. For doing that, we just apply Uniform Weighting (see Section~\ref{sec:weight}) to the graph in Algorithm~\ref{algo:location}.
 \item \textbf{Intra-Cluster (IC):} Uses the geographic similarity defined in Section~\ref{sec:weight} for defining intra-cluster similarities between users. The idea is as follows. The weight of any edge $(v,v')\in E$ is now calculated by summing up $w^{\text{geo}}(v,v')$ for all clusters where $u_v$ and $u_{v'}$ have photos, and dividing the resulting sum by the total number of clusters. After that, we again just need to plug the weighted graph into Algorithm~\ref{algo:location} for computing recommendations.
 \item \textbf{Collaborative Filtering (CF):} This is the geographic version of the classic user-based collaborative filtering algorithm. We use the correlation weighting scheme defined in Section~\ref{sec:weight} where profile vectors' components are the geographic clusters where users took pictures, within and out of the context of interest. 
 \item \textbf{Two-Layer (TL):} This is the algorithm originally introduced in~\cite{matyas2009}. We weigh the relational graph with the \textit{two-layer similarity} and plug it into the recommendation algorithm. 
 \item \textbf{Correlation + Two-Layer (CF-TL)}: We found empirically that TL works best for cold-start scenarios, while CF outperforms the other algorithms for non cold-start scenarios (see Figures~\ref{fig:some_out} and ~\ref{fig:all_out}). This gave us the insight to propose a weighting combination strategy where we weigh each relation differently according to the case presented. If the target user is a cold-start user, we weigh his relations with the \textit{two-layer similarity}, if not, we use the \textit{correlation weighting}. The flexibility for combining different weighting schemes in such an easy way is one of the main advantages of our approach.
\end{itemize}

\paragraph{Results}

Figure~\ref{fig:some_out} depicts the results for the \textit{leave-some-out} scenario. Notice that when there is enough data available, CF outperforms all the other methods. The Intra-Cluster recommender, although worse than CF and TL, is better than MP in all cases. This indicates that the geographic similarity indeed is able to capture some preferences of the user, under the assumption that users tend to like the items of other users that are nearby the items they have selected in the past.

Figure~\ref{fig:all_out} shows the results for the \textit{leave-all-out}. Notice that TL is the winner in this scenario. This is in line with the results of~\cite{matyas2009}, where they showed that geographic partonomies can help to improve the recommendations in cold-start scenarios. Notice that since we remove all the photos of all test users in the context of interest, we just have one possible split of training/test, and thus cannot compute standard deviations and plot error bars.

\begin{figure}[htp]
\vspace{-0.2cm}
\centering
\includegraphics[width=0.99\linewidth]{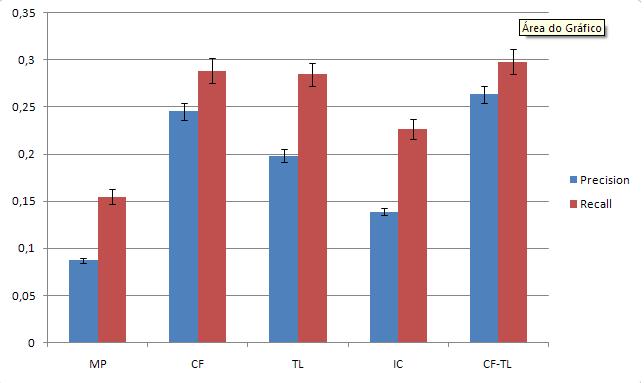}
\caption{Evaluation on the \textit{leave-some/all-out} scenario when half of the users are cold-start users.}\label{fig:mix_05}
\end{figure}

\begin{figure}
\centering
\includegraphics[width=0.99\linewidth]{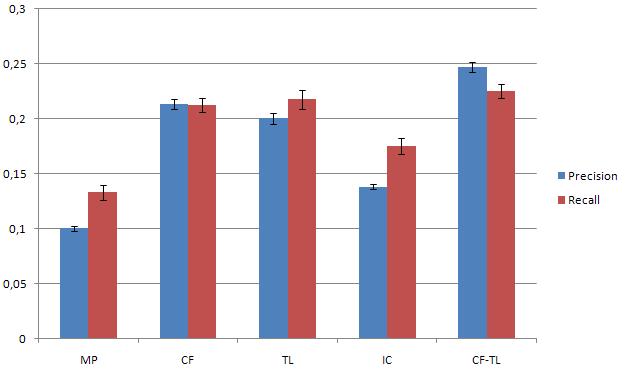}
\caption{Evaluation on the \textit{leave-some/all-out} scenario when 70\% of the users are cold-start users.}\label{fig:mix_07}
\end{figure}

In Figure~\ref{fig:mix_05} we show the results of the \textit{leave-some/all-out} scenario when 50\% of the test users are cold-start users. In this case, the combination method CF-TL is slightly better, both in precision and recall, than the other methods. But when 70\% of ther users are cold-start users (see Figure~\ref{fig:mix_07}) the superiority of CF-TL becomes more evident. We also evaluated this scenario when 30\% of the users are cold-start users, but since there was no significant differences in performance in comparison to 50\% of cold-start users, we do not show those results here. 

\subsection{HP ePrint Experiments}
The HP ePrint Mobile Print Location (MPL)~\footnote{http://www.hp.com/go/eprintmobile} service allows smartphone 
users to print documents, photos, web pages and emails directly to public print providers, such as business centers at hotels and airports, 
and dedicated copy and print service stores. Users can query for providers near their current location and then submit a print job upon 
which a pickup code is generated. This code can then be used in the store to obtain the printout. 

\paragraph{Data Preparation} Our dataset contains a 5-month trace of all print jobs sent through the service. To protect business sensitive information the users as well as providers were anonymized. Furthermore, the volume was obfuscated using a bootstrap inspired sampling algorithm that takes uniform random samples with replacement of print jobs from the original dataset and replays them with the original timestamp randomly modified up to one week from the original job submission. We sampled 100k print jobs in this way, and then extracted all the entries that occurred in Manhattan (about 2,580). Each entry comprises the user id of the user submitting the print job, the provider id of the provider receiving the print job, the timestamp of the print job and the latitude and longitude of the print provider location. 

\paragraph{Evaluation Protocol} Here the geographic items are print providers and in contrast to Panoramio, the POIs are well defined, and therefore we do not need to compute geographic clusters. We chose Manhattan as the geographic context of interest because of its large volume of printing activity.

Here we used the \textit{leave-one-out} protocol, i.e., for each user we randomly removed one of the items and used the rest for computing recommendations. 
We considered all the users who have used at least one print provider, i.e., all users in Manhattan, for our test set. 
due to the nature of the data set, we only evaluate the \textit{leave-some/all-out} scenario, where some users are cold-start users and others not. 
Furthermore, we ignore users for which the recommendation list is empty. Since we only hide one photo, we do not compute precision, 
because this would be the same as recall up to a multiplicative constant. 
Thus, we only compute the average recall over ten random splits of training/test. 
Again, due to the nature of this dataset we only evaluated the MP and CF algorithms for top-$10$ recommendations. 
Whenever CF is not able to fill the recommendation list up to 10, we fill up the list with the most popular items that are not already in the recommended list.

\begin{figure}
\centering
\includegraphics[width=0.99\linewidth]{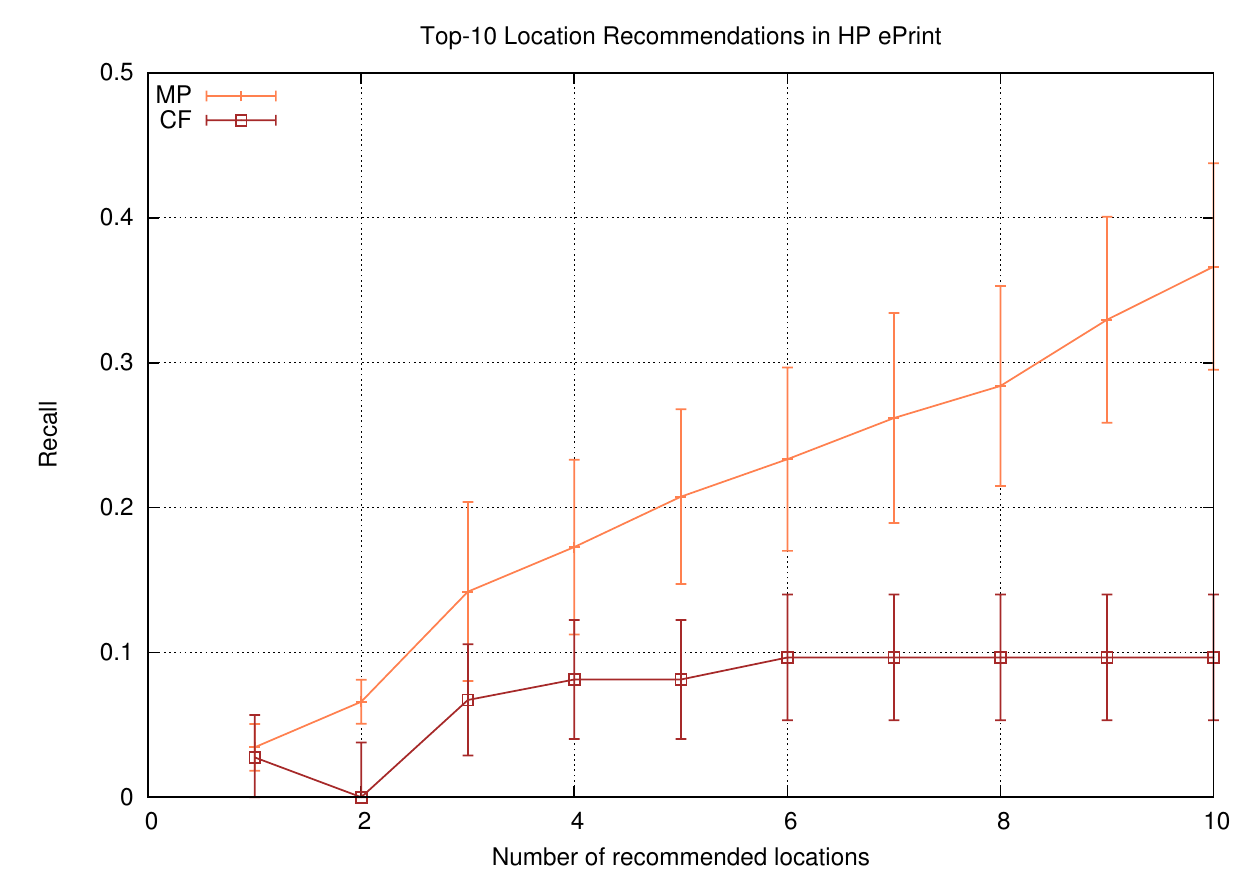}
\caption{MP vs. CF in HP ePrint.}\label{fig:eprint_results}
\end{figure}

\paragraph{Results}
Figure~\ref{fig:eprint_results} shows the results for MP and CF in ePrint. The error bars represent one standard deviation.
We can see that the most popular item algorithm outperforms
the collaborative filtering algorithm consistently, in particular for long recommendation lists. Given the relative short period this service
has been on the market it is dangerous to draw too many conclusions from this behavior, but there are two lessons to be learned from this result.
First, we could easily adapt our algorithm to generate useful recommendations for a use case vastly different than the Panoramio application,
by just simply applying a different weighting scheme. Second, a very simple
popularity-based recommendation engine can do very well, and deploying sophisticated collaborative filtering techniques
may not always be the best choice.
 
\section{Related Work}
\label{sec:related_work}
Related work falls into three broad categories: relational classification and similarity measures,
location-aware methods, and methods focusing on improving the computational scalability of the
model building and execution phases.

\paragraph{Relational Classification and Similarity Measures} Relational classification has been applied to areas where entities are linked in an explicit manner, like hypertext documents, such that the class of a target instance only depends on the class of its related instances~\cite{preisach06ensemble}. In~\cite{marinho09tag} we presented a relational neighbor classification approach in the ECML/PKDD Discovery Challenge 2009, which was about tag recommendations. Similarly to that work, in this paper we formulate the recommendation problem in a relational neighbor framework where geographic contexts link entities represented by user/location pairs.

In~\cite{matyas2009} a new similarity measure for computing location recommendations based on a non-overlapping hierarchical taxonomy of locations is presented. The key idea is that co-activity in locations can be better captured if you zoom out to larger and larger locations. Similar to our work they use a panoramio data set to evaluate their algorithm. However, their model assumes a cold-start user is making the query, i.e. a user who has no trace in the geographic context of the query. The reliance on a place taxonomy is also more restricting than our general model. 

Jeh and Widom define SimRank in~\cite{jeh2002}. The SimRank similarity measure is as the name suggests highly influenced by the 
more famous PageRank~\cite{brin1998} algorithm. The general idea is that ``two objects are similar if they are referenced by similar objects''~\cite{jeh2002}.
This definition recursively propagates similarity through a relational graph to leverage structural context in addition to the more
traditional use of object content and attribute information. In our work akin to the model used in~\cite{matyas2009}, we also have a recursive, relation-based definition of similarity but in the form of hierarchical geographic contexts. For faster
computation we simplify the similarity model to not be recursive between leaf-nodes within the same geographic context in our graph. Our evaluations show that this can be done without any loss of accuracy if the user who requested the recommendation already has 
an activity trace within the geographic context for which the recommendation is sought.   
Minkov and Cohen~\cite{minkov2010} also explore relational properties among objects
to compute similarity scores. They focus on the problem of effectively searching in graphs comprised of interrelated objects of various types. They propose effective random-walk based procedures to evaluate the Personal PageRank measure~\cite{brin1998}, 
a measure that takes into account the scope of the query in addition to the well-known random surfer assumption of the
original PageRank algorithm. 

\paragraph{Location-aware Methods} In~\cite{ye2011} a standard user-based collaborative filtering approach is extended with a novel mechanism of geographic influence based on a statistical model of how likely a user is to check-in to two places based on their distance. The geographic model is a distributional assumption (power law) fitted to real data sets from two popular check-in services.

A number of studies have looked at timestamped GPS traces to predict future locations within very restricted geographic regions~\cite{cao2010,zheng2010b,burbey2008}. Markov models and tensor factorization models are fit to the data and non-personalized predictions of the most likely next location or the most likely activity given a location and time are produced. There is a lot of novel work on automatically detecting geographic context in these papers but the general approach cannot be replicated easily in our scenario since we do not have the same luxury of rich traces as we mainly focus on implicit feedback as input. Relying on GPS traces is furthermore a privacy concern and has scalability and power consumption implications. We produce more personalized recommendations given user-user similarities as opposed to just looking at the most popular or most frequent behavior. Furthermore large Markov models and tensor factorization algorithms tend to be very costly to compute, which would therefore need to be done in an off-line setting whereas we also target real-time recommendations.

Bayesian networks have also been studied to model and learn patterns in location, time and weather contexts for individual users in~\cite{park2007}. However, compared to our model these models tend to be complex and require expert human knowledge to construct, and furthermore they are not tractable.  ~\cite{brunato2003} applies a center of mass model to detect and recommend locations and POIs. This work was before the check-in systems era so now we could just more easily query the check-in services for this information.
In~\cite{horozov2006} user-user based CF with location pre-filtering is employed in an explicit voting scenario. The cold start problem is solved by generating random recommendations using pseudo users. We address the problem by incorporating out-of geo context similarities for in-context recommendations which is less ad-hoc.

In our previous work in~\cite{sandholm2010} we studied popularity inferred recommendations based on location, friends and tag prefiltering using both explicit and implicit feedback. In this work we extend that model to provide personalized recommendations similar to the work in~\cite{sandholm2011} but with a new model that incorporates location, and distance metrics directly in the evaluation graph as opposed to relying on ad-hoc and costly pre and post filtering.

The GeoFolk system~\cite{sizov2010} was designed to take both geographic context and text features into account for various
information retrieval tasks such as tag recommendation, content classification and clustering. Experiments show that combining
both textual and geographic relevance leads to more accurate results than using the two factors in isolation. Although our methods
and use case targets are quite different from this work, the empirical evidence of the influence geographic context has on
information retrieval is promising and serves as motivation for our work. 

\paragraph{Computational Scalability} Popular and accurate recommender system methods such as those based on matrix factorization can incure very high model building as well as execution overhead, in particular as more contextual factors beyond users and items are taken into account. As a result there have been many attempts at improving the computational scalability of pre-existing methods.

In~\cite{sandholm2011,ye2010,prete2010} the general issue of complex and high-latency model building and execution for location-based recommendations is addressed. To achieve real-time performance~\cite{sandholm2011} and~\cite{prete2010} pre-filter based on location and~\cite{ye2010} pre-filters based on friends to reduce the complexity of the models. As opposed to pre- or post-filtering context we make contextual paramaters an inherent part of our graph model to allow interesting combinations of various types of context in an efficient way.

In~\cite{rendle2011} a fast context-aware recommendation algorithm is proposed that maintains the features of state-of-the-art multi-tensor matrix factorization while bringing down the complexity of the previously known algorithms from exponential to linear growth in problem size. The main idea is to solve the least-squares optimization problem for each model parameter separately. Our method in contrast achieves scalability by not utilizing any complex matrix factorization.

\section{Conclusions and Future Work}
\label{sec:conclusions}
In this paper we introduced a relational graph for capturing the geographic preferences of users with the purpose of
generating personalized recommendations in services with geotagged content. 
We also presented several weighting schemes for representing different types of user preferences in the 
proposed graph. Furthermore, we propose a recommendation algorithm template that is 
sufficiently generic to derive many traditional and new location-aware recommendation algorithms, 
including combinations of them, by only requiring changes in the definition 
of the edge weights. Assuming the graph is given, the algorithm requires modest computational effort since it runs linearly in the 
number of neighbors. We have tested the proposed algorithms with two real-world datasets, geotagged photos from Panoramio and 
print jobs from the HP ePrint Mobile Print Location service, and showed how our model easily suits many different 
recommendation scenarios. 

We also gained insights about which notion of similarity works best for a set of scenarios. In cold-start scenarios a 
geographic partonomy seems to be a good alternative, whereas when there is enough data available the plain location-aware 
collaborative filtering algorithm yields the best result. 
In response to this finding, we proposed to combine a partonomy-based similarity measure with the cosine similarity 
by weighting individual relations in the graph according to the type of the user, i.e., cold-start versus non cold-start. 
By doing this, we achieved better recall and precision in particular in scenarios where there are many cold-start users.

As future work, we plan to incorporate temporal aspects in the model, such that the items to be 
recommended match the temporal context of the user. For example, it may not make as much sense to 
recommend ski resorts in New York during the summer as it would to make the same recommendation during the peak winter season. 
Another natural extension of our work would be to assign the weights in our graph based on the strength of the
social ties between the users, e.g. based on their declared or implied social networks.
Finally, we plan to investigate machine learning approaches for learning optimal weights based on the location-aware 
recommendation task at hand.

 \paragraph{Acknowledgements} This work was supported by a cooperation with Hewlett-Packard Brasil Ltda. using incentives of Brazilian Informatics Law (Law No. 8.2.48 of 1991). We would also like to thank Christina Aperjis, Sitaram Asur and Mao Ye for insightful comments on our work.

\bibliographystyle{abbrv}


\begin{thebibliography}{10}

\bibitem{brin1998}
S.~Brin and L.~Page.
\newblock The anatomy of a large-scale hypertextual web search engine.
\newblock In {\em Proceedings of the seventh international conference on World
  Wide Web 7}, WWW7, pages 107--117, Amsterdam, The Netherlands, The
  Netherlands, 1998. Elsevier Science Publishers B. V.

\bibitem{brunato2003}
M.~Brunato and R.~Battiti.
\newblock Pilgrim: A location broker and mobility-aware recommendation system.
\newblock In {\em PERCOM '03: Proc. of the First IEEE International Conference
  on Pervasive Computing and Communications}, pages 265--. IEEE Computer
  Society, 2003.

\bibitem{burbey2008}
I.~Burbey and T.~L. Martin.
\newblock Predicting future locations using prediction-by-partial-match.
\newblock In {\em MELT '08: Proc. of the first ACM international workshop on
  Mobile entity localization and tracking in GPS-less environments}, pages
  1--6. ACM, 2008.

\bibitem{cao2010}
X.~Cao, G.~Cong, and C.~S. Jensen.
\newblock Mining significant semantic locations from gps data.
\newblock {\em Proc. VLDB Endow.}, 3:1009--1020, September 2010.

\bibitem{chak98}
S.~Chakrabarti, B.~Dom, and P.~Indyk.
\newblock Enhanced hypertext categorization using hyperlinks.
\newblock {\em SIGMOD Rec.}, 27(2):307--318, 1998.

\bibitem{prete2010}
L.~Del~Prete and L.~Capra.
\newblock differs: A mobile recommender service.
\newblock In {\em MDM '10: Proc. of the 2010 Eleventh International Conference
  on Mobile Data Management}, pages 21--26. IEEE Computer Society, 2010.

\bibitem{Ester96adensity-based}
M.~Ester, H.~peter Kriegel, J.~S, and X.~Xu.
\newblock A density-based algorithm for discovering clusters in large spatial
  databases with noise.
\newblock pages 226--231. AAAI Press, 1996.

\bibitem{horozov2006}
T.~Horozov, N.~Narasimhan, and V.~Vasudevan.
\newblock Using location for personalized poi recommendations in mobile
  environments.
\newblock In {\em Proc. of the International Symposium on Applications on
  Internet}, pages 124--129. IEEE Computer Society, 2006.

\bibitem{jeh2002}
G.~Jeh and J.~Widom.
\newblock Simrank: a measure of structural-context similarity.
\newblock In {\em Proceedings of the eighth ACM SIGKDD international conference
  on Knowledge discovery and data mining}, KDD '02, pages 538--543, New York,
  NY, USA, 2002. ACM.

\bibitem{Lu03link-basedclassification}
Q.~Lu and L.~Getoor.
\newblock Link-based classification using labeled and unlabeled data.
\newblock In {\em Proceedings of the ICML Workshop on The Continuum from
  Labeled to Unlabeled Data in Machine Learning and Data Mining}, 2003.

\bibitem{Macskassy03asimple}
S.~A. Macskassy and F.~Provost.
\newblock A simple relational classifier.
\newblock In {\em MRDM '03: Proceedings of the Second Workshop on
  Multi-Relational Data Mining at KDD-2003}, pages 64--76, 2003.

\bibitem{marinho09tag}
L.~B. Marinho, C.~Preisach, and L.~Schmidt-Thieme.
\newblock Relational classification for personalized tag recommendations.
\newblock In {\em Proc. of the ECML/PKDD Discovery Challenge}, volume 497 of
  {\em CEUR-WS.org}, 2009.

\bibitem{matyas2009}
C.~Matyas and C.~Schlieder.
\newblock A spatial user similarity measure for geographic recommender systems.
\newblock In K.~Janowicz, M.~Raubal, and S.~Levashkin, editors, {\em GeoSpatial
  Semantics}, volume 5892 of {\em Lecture Notes in Computer Science}, pages
  122--139. Springer Berlin / Heidelberg, 2009.

\bibitem{minkov2010}
E.~Minkov and W.~W. Cohen.
\newblock Improving graph-walk-based similarity with reranking: Case studies
  for personal information management.
\newblock {\em ACM Trans. Inf. Syst.}, 29:4:1--4:52, December 2010.

\bibitem{park2007}
M.-H. Park, J.-H. Hong, and S.-B. Cho.
\newblock {Location-Based Recommendation System Using Bayesian User's
  Preference Model in Mobile Devices}.
\newblock pages 1130--1139. 2007.

\bibitem{preisach06ensemble}
C.~Preisach and L.~Schmidt-Thieme.
\newblock Relational ensemble classification.
\newblock In {\em ICDM '06: Proc. of the 6th International Conference on Data
  Mining}, pages 499--509. IEEE Computer Society, 2006.

\bibitem{rendle2011}
S.~Rendle, Z.~Gantner, C.~Freudenthaler, and L.~Schmidt-Thieme.
\newblock Fast context-aware recommendations with factorization machines.
\newblock In {\em Proceedings of the 34th international ACM SIGIR conference on
  Research and development in Information}, SIGIR '11, pages 635--644, New
  York, NY, USA, 2011. ACM.

\bibitem{sandholm2011}
T.~Sandholm and H.~Ung.
\newblock Real-time location-aware collaborative filtering of web content.
\newblock In {\em CaRR '11: Proc. of the 2011 Workshop on Context-awareness in
  Retrieval and Recommendation}, pages 14--18. ACM, 2011.

\bibitem{sandholm2010}
T.~Sandholm, H.~Ung, C.~Aperjis, and B.~A. Huberman.
\newblock Global budgets for local recommendations.
\newblock In {\em RecSys '10: Proc. of the fourth ACM conference on Recommender
  systems}, pages 13--20. ACM, 2010.

\bibitem{sizov2010}
S.~Sizov.
\newblock Geofolk: latent spatial semantics in web 2.0 social media.
\newblock In {\em Proceedings of the third ACM international conference on Web
  search and data mining}, WSDM '10, pages 281--290, New York, NY, USA, 2010.
  ACM.

\bibitem{ye2011}
M.~Ye, P.~Yi, W.-C. Lee, and D.-L. Lee.
\newblock Exploiting geographical influence for collaborative
  point-of-interests recommendation.
\newblock In {\em SIGIR '11: Proc. of the 34th ACM SIGIR Conference}. ACM,
  2011.
\newblock (to appear).

\bibitem{ye2010}
M.~Ye, P.~Yin, and W.~C. Lee.
\newblock {Location recommendation for location-based social networks}.
\newblock In {\em Proc. of the 18th SIGSPATIAL International Conference on
  Advances in Geographic Information Systems}, GIS '10, pages 458--461. ACM,
  2010.

\bibitem{zheng2010b}
V.~W. Zheng, Y.~Zheng, X.~Xie, and Q.~Yang.
\newblock {Collaborative location and activity recommendations with GPS history
  data}.
\newblock In {\em WWW'10: Proc. of the 19th international conference on World
  Wide Web}, pages 1029--1038. ACM, 2010.

\end{thebibliography}
%
%
\balancecolumns
\end{document}